\documentclass[aps,twocolumn,a4paper,10pt]{revtex4-1}
\usepackage{mathrsfs}
\usepackage{graphicx}
\usepackage{latexsym}
\usepackage{amsmath}
\usepackage{amssymb}
\usepackage{textcomp}
\usepackage{amsbsy}
\usepackage{color}
\usepackage{epstopdf}

\begin{document}
\title{\large \bf Radion tunneling in modified theories of gravity}
\author{Tanmoy Paul }
\email{pul.tnmy9@gmail.com}
\author{Soumitra SenGupta}
\email{tpssg@iacs.res.in}
\affiliation{Department of Theoretical Physics,\\
Indian Association for the Cultivation of Science,\\
2A $\&$ 2B Raja S.C. Mullick Road,\\
Kolkata - 700 032, India.\\}

\begin{abstract}
We consider a five dimensional warped spacetime where the bulk geometry is governed by higher curvature 
$F(R)$ gravity. In this model, we determine the modulus potential originating from the scalar degree of freedom of higher 
curvature gravity. In the presence of this potential, we investigate the possibility of 
modulus (radion) tunneling leading to an instability in the brane configuration. Our results 
reveal that the parametric regions where the tunneling probability is highly suppressed, corresponds to 
the parametric values required to resolve the gauge hierarchy problem.
\end{abstract}
\maketitle

\section{Introduction}
Over the last two decades models with extra spatial dimensions \cite{arkani,horava,RS,kaloper,cohen,burgess,chodos}  
have been increasingly playing a central role in search for physics beyond standard model of elementary particle \cite{rattazzi,tp_fermion} 
and Cosmology \cite{marteens,tp3}. Such higher dimensional 
scenarios occur naturally in string theory and also are viable candidates to resolve the well known gauge hierarchy problem. 
Depending on different possible compactification schemes 
for the extra dimensions, a large number of models have been constructed. In all these models, 
our visible universe is identified as a 3-brane embedded in a higher dimensional spacetime 
and  is described  through  a low energy effective theory on the brane carrying the signatures 
of extra dimensions \cite{kanno,shiromizu,sumanta}.\\
Among various extra dimensional models proposed over last several years, warped 
extra dimensional model pioneered by Randall and Sundrum (RS) \cite{RS} earned a 
special attention since it resolves the gauge hierarchy problem without introducing any 
intermediate scale (between Planck and TeV) in the theory. Subsequently different variants 
of warped geometry model were extensively studied in 
\cite{GW,GW_radion,csaki,julien,ssg1,ssg2,tp1,tp2,tp3,sumanta_cosmology,anjan}. A generic feature of many of these models is that 
the bulk spacetime is endowed with high curvature scale $\sim$ 4 dimensional Planck scale.\\
It is well known that Einstein-Hilbert action can be generalized
by adding higher order curvature terms 
which naturally arise from diffeomorphism property of the action. 
Such terms also have their origin in String theory from  
quantum corrections. In this context $F(R)$ \cite{1,2,3,4,5,6,7,8,9,10,faraoni,felice,paliathanasis}, Gauss-Bonnet (GB) \cite{nojiri2,tp-nb,cognola}
or more generally Lanczos-Lovelock gravity 
are some of the candidates in higher curvature gravitational 
theory.\\ 
In general the higher curvature terms are suppressed with respect to Einstein-Hilbert term 
by Planck scale. Hence in low curvature regime, their contributions are negligible. However 
higher curvature terms become extremely relevant in a region with 
large curvature. Thus for bulk geometry where the curvature is of the  
order of Planck scale, the higher curvature terms should play 
a crucial role. Motivated by this idea, in the present work, we consider 
a generalized 
warped geometry model by replacing Einstein-Hilbert bulk gravity action with a higher 
curvature $F(R)$ gravitational theory \cite{faraoni,felice,barrow,marino,bahamonde,catena,tp_higher,sumanta_higher,sumanta_lhc}.\\
One of the crucial aspects of higher dimensional two brane models is to stabilize the interbrane separation (also known as modulus 
or radion). For this purpose, one needs to generate a suitable radion potential with a stable minimum \cite{GW,GW_radion,csaki}. The presence 
of such minimum guarantees the stability of the modulus field. In Goldberger-Wise stabilization mechanism \cite{GW,GW_radion}, an 
external bulk scalar field was invoked to create such a stable radion potential. However, when the bulk is endowed with higher curvature F(R) gravity, 
then apart from the metric there is an additional scalar degree of freedom originating from higher derivative terms of the metric. It 
has been shown that such a scalar degree of freedom can play the role of the stabilizing field for appropriate choices 
of the underlying $F(R)$ model \cite{ssg2,tp1}.\\ 
It is important to analyze the exact nature of the resulting radion potential to explore whether there exists 
a metastable minimum for the radion from which it can tunnel and leads to an instability of the braneworld 
\cite{giddings,pilado,coleman1,coleman2,parke}. In this paper, we aim to determine the radion tunneling in the 
presence of higher curvature gravity in the bulk.\\
Our paper is organized as follows : Following section is devoted to brief review of the conformal relationship between $F(R)$ and scalar-tensor
(ST) theory. In section III, we extend our analysis of
section II for the specific $F(R)$ model considered in this work. Section IV extensively describes the tunneling probability for the dual ST model
while section V addresses these for the original $F(R)$ model. After discussing the equivalence, the paper ends 
with some concluding remarks in section VI.

\section{Transformation of a F(R) theory to scalar-tensor theory}

In this section, we briefly describe how a higher curvature F(R) gravity model in five dimensional 
scenario can be recast into Einstein gravity with a scalar field. 
The F(R) action is expressed as,
\begin{equation}
 S = \frac{1}{2\kappa^2}\int d^4x d\phi \sqrt{G} F(R)
 \label{action0}
\end{equation}
where $x^{\mu} = (x^0, x^1, x^2, x^3)$ are usual four dimensional coordinate and 
$\phi$ is the extra dimensional spatial angular coordinate. 
$R$ is the five dimensional Ricci curvature and $G$ is the determinant of the metric. 
Moreover $\frac{1}{2\kappa^2}$ is taken as $2M^3$ where $M$ is the five dimensional 
Planck scale. Introducing an auxiliary field $A(x,\phi)$, 
the action (in eqn.(\ref{action0})) can be equivalently written as,
\begin{equation}
 S = \frac{1}{2\kappa^2}\int d^4x d\phi \sqrt{G} [F'(A)(R-A) + F(A)]
 \label{action00}
\end{equation}
By the variation of the auxiliary field $A(x,\phi)$, one easily obtains $A=R$. 
Plugging back this solution $A=R$ into action (\ref{action00}), 
initial action (\ref{action0}) can be reproduced. At this stage, one may perform a 
conformal transformation of the metric as
\begin{equation}
  G_{MN}(x,\phi) \rightarrow \tilde{G}_{MN} = e^{\sigma(x,\phi)}G_{MN}(x,\phi)
 \nonumber\\
\end{equation}
$M, N$ run form 0 to 5. $\sigma(x,\phi)$ is conformal factor and related to the auxiliary 
field as $\sigma = (2/3)\ln F'(A)$. Using this 
relation between $\sigma(x,\phi)$ and $A(x,\phi)$, one lands up with the following scalar-tensor action
\begin{eqnarray}
 S = \frac{1}{2\kappa^2}&\int& d^4x d\phi \sqrt{\tilde{G}} \bigg[\tilde{R} + 3\tilde{G}^{MN}\partial_M\sigma \partial_N\sigma \nonumber\\
 &+&4\tilde{G}^{MN}\partial_M\partial_N\sigma - \bigg(\frac{A}{F'(A)^{2/3}} - \frac{F(A)}{F'(A)^{5/3}}\bigg)\bigg]
 \nonumber\\
\end{eqnarray}
where $\tilde{R}$ is the Ricci scalar formed by $\tilde{G}_{MN}$. $\sigma(x,\phi)$ is 
the scalar field emerged from 
higher curvature degrees of freedom. Clearly kinetic part of $\sigma(x,\phi)$ is
non canonical. In order to make the scalar field canonical, transform 
$\sigma$ $\rightarrow$ $\Psi(x,\phi) = \sqrt{3}\frac{\sigma(x,\phi)}{\kappa}$. In terms of 
$\Psi(x,\phi)$, the above action takes 
the form,
\begin{eqnarray}
 S = \int d^4x d\phi \sqrt{\tilde{G}} &\bigg[&\frac{\tilde{R}}{2\kappa^2} + \frac{1}{2}\tilde{G}^{MN}\partial_M\Psi\partial_N\Psi\nonumber\\ 
 &+&\frac{2}{\sqrt{3}\kappa}\tilde{G}^{MN}\partial_M\partial_N\Psi - U(\Psi)\bigg]
 \nonumber\\
\end{eqnarray}

where $U(\Phi) = \frac{1}{2\kappa^2}[\frac{A}{F'(A)^{2/3}} - \frac{F(A)}{F'(A)^{5/3}}]$ is 
the scalar field potential which depends on the 
form of $F(R)$. Thus the action of $F(R)$ gravity in five dimension can be transformed 
into the action of a scalar-tensor 
theory by a conformal transformation of the metric.

\section{Warped spacetime in F(R) model and corresponding scalar-tensor theory}

In the present paper, we consider a five dimensional spacetime with two 3-brane scenario in F(R) model. The form of 
$F(R)$ is taken as $F(R) = R + \alpha R^n$ where $n$ takes only positive values, $\alpha$ is a constant and has the 
mass dimension $[2-2n]$. Considering $\phi$ as the extra dimensional 
angular coordinate, two branes are located at $\phi = 0$ (hidden brane) and at $\phi = \pi$ (visible brane) respectively 
while the latter one is identified with the visible universe. Moreover the spacetime is $S^1/Z_2$ 
orbifolded along the coordinate $\phi$. The action for this model is :
\begin{eqnarray}
 S&=&\int d^4x d\phi \sqrt{G} \bigg[\frac{1}{2\kappa^2}\bigg(R + \alpha R^n\bigg)\nonumber\\ 
 &+&\frac{1}{r_c}\bigg(V_h+\frac{Q_h}{\kappa^2}\bigg)\delta(\phi) + \frac{1}{r_c}\bigg(V_v+\frac{Q_v}{\kappa^2}\bigg)\delta(\phi-\pi)\bigg]
 \label{actionF(R)}
\end{eqnarray}
where $V_h$, $V_v$ are the brane tensions on hidden, visible brane 
respectively. We also include Gibbons-Hawking boundary terms on the branes, symbolized by $Q_h$ and $Q_v$ in the above action 
(i.e $Q_h$, $Q_v$ are the trace of extrinsic curvatures on hidden, visible brane respectively).

This higher curvature $F(R)$ model (in eqn.(\ref{actionF(R)})) can be transformed into a scalar-tensor 
theory by using the technique discussed in the previous section. Performing a conformal transformation of the metric as 
\begin{equation}
 G_{MN}(x,\phi) \rightarrow \tilde{G}_{MN} = \exp{(\frac{1}{\sqrt{3}}\kappa\Psi(x,\phi))}G_{MN}(x,\phi)
 \label{conformal}
\end{equation}
the above action (in eqn.(\ref{actionF(R)})) can be expressed as a scalar-tensor theory with the action given by :
\begin{eqnarray}
 S&=&\int d^4x d\phi \sqrt{\tilde{G}} \bigg[\frac{\tilde{R}}{2\kappa^2} + \frac{1}{2}\tilde{G}^{MN}\partial_M\Psi \partial_N\Psi - U(\Psi)\nonumber\\
 &+&\Lambda + \frac{2}{\sqrt{3}\kappa}\tilde{G}^{MN}\partial_M\partial_N\Psi 
 + \frac{1}{r_c}e^{-\frac{5}{2\sqrt{3}}\kappa\Psi} \bigg(V_h+\frac{Q_h}{\kappa^2}\bigg)\delta(\phi)\nonumber\\ 
 &+&\frac{1}{r_c}e^{-\frac{5}{2\sqrt{3}}\kappa\Psi} \bigg(V_v+\frac{Q_v}{\kappa^2}\bigg)\delta(\phi-\pi)\bigg]
 \label{action1ST}
\end{eqnarray}
where $\Lambda$ is chosen to be negative and the quantities in tilde are reserved for ST theory. $\tilde{R}$ is the Ricci curvature formed 
by the transformed metric $\tilde{G}_{MN}$. $\Psi(x,\phi)$ is the scalar 
field corresponds to higher curvature degree of freedom and $U(\Psi)$ is the scalar potential which 
for this specific form of $F(R)$ has the form,
\begin{eqnarray}
 U(\Psi) = \Lambda + \frac{a}{\kappa^2}e^{b\kappa \Psi}
 \label{scalar_potential}
\end{eqnarray}

where $a$ (mass dimension [2]) and $b$ (dimensionless) are related to the parameters 
$\alpha$ and $n$ by the following expressions :
\begin{eqnarray}
 \alpha&=&\bigg(\frac{2\sqrt{3}a^2\kappa^{8/3}}{4b+10}\bigg) \bigg[\frac{\sqrt{3}a\kappa^{8/3}}{4b^2}\bigg]^{\frac{3}{2+2\sqrt{3}b}}\nonumber\\
 n&=&\frac{5 + 2\sqrt{3}b}{2 + 2\sqrt{3}b}
 \label{a and b}
 \end{eqnarray}
 
 Considering that the scalar field depends on extra dimensional coordinate only (see eqn.(\ref{sol.scalar.field})), the 
 total derivative term can be integrated once leading to the final form of the action as follows:
 
 \begin{eqnarray}
  S&=&\int d^4x d\phi \sqrt{\tilde{G}} \bigg[\frac{\tilde{R}}{2\kappa^2} + \frac{1}{2}\tilde{G}^{MN}\partial_M\Psi \partial_N\Psi - U(\Psi)\nonumber\\
 &+&\Lambda + \bigg(\frac{1}{r_c}e^{-\frac{5}{2\sqrt{3}}\kappa\Psi} \big(V_h+\frac{Q_h}{\kappa^2}\big) 
 + \frac{2}{\sqrt{3}\kappa}\tilde{G}^{\phi\phi}\frac{\partial\Psi}{\partial\phi}\bigg)\delta(\phi)\nonumber\\ 
 &+&\bigg(\frac{1}{r_c}e^{-\frac{5}{2\sqrt{3}}\kappa\Psi} \big(V_v+\frac{Q_v}{\kappa^2}\big) 
 + \frac{2}{\sqrt{3}\kappa}\tilde{G}^{\phi\phi}\frac{\partial\Psi}{\partial\phi}\bigg)\nonumber\\
 &\delta&(\phi-\pi)\bigg]
 \label{action_newST}
 \end{eqnarray}

 \section{Radion potential and tunneling probability in scalar-tensor (ST) theory}
In order to generate radion potential in ST theory, here we adopt the GW mechanism \cite{GW_radion} which requires 
a scalar field in the bulk. For the case of ST theory presented in eqn.(\ref{action_newST}), $\Psi$ can act as a bulk 
scalar field. Considering a negligible backreaction 
of the scalar field ($\Psi$) on the background spacetime, the solution of metric $\tilde{G}_{MN}$ 
is exactly same as well known RS model i.e.
\begin{equation}
 d\tilde{s}^2 = e^{- 2 kr_c|\phi|} \eta_{\mu\nu} dx^{\mu} dx^{\nu} - r_c^2d\phi^2
 \label{grav.sol1.ST}
\end{equation}
where $k = \sqrt{\frac{-\Lambda}{24M^3}}$. Using these metric and explicit form of $U(\Psi)$, 
we obtain the scalar field equation of motion 
in the bulk as follows,
\begin{eqnarray}
 \frac{1}{r_c^2}\frac{\partial^2\Psi}{\partial\phi^2} - 4\frac{k}{r_c}\frac{\partial\Psi}{\partial\phi} + \frac{ab}{\kappa}e^{b\kappa\Psi} = 0
 \label{eom.scalar.field}
\end{eqnarray}
To derive the above equation of motion, $\Psi$ is taken as function of extra dimensional coordinate only. Considering 
the variation of $\Psi(\phi)$ is small in the bulk \cite{GW,GW_radion}, eqn.(\ref{eom.scalar.field}) turns out to be,
\begin{eqnarray}
 - 4\frac{k}{r_c}\frac{\partial\Psi}{\partial\phi} + \frac{ab}{\kappa}e^{b\kappa\Psi} = 0
 \label{modified.eom.scalar.field}
\end{eqnarray}
With a non zero value of $\Psi$ on the branes, above equation has the following solution :
\begin{equation}
 e^{b\kappa\Psi(\phi)} = \frac{4k}{ab^2} \bigg[\frac{1}{y_0 - r_c\phi}\bigg]
 \label{sol.scalar.field}
\end{equation}
where $y_0 = \frac{4k}{ab^2}e^{-b\kappa v_h}$ and $v_h$ is the value of the bulk scalar field on the 
hidden brane ($\phi=0$).\\
Using the solution of metric (see eqn.(\ref{grav.sol1.ST})), we obtain the extrinsic curvature 
of $\phi=$ constant hypersurface as follows :
\begin{eqnarray}
 Q_{\mu\nu} = ke^{-2kr_c\phi} \eta_{\mu\nu}
 \label{extrinsic1}
\end{eqnarray}
and
\begin{eqnarray}
 Q = Q_{\mu\nu} e^{2kr_c\phi} \eta^{\mu\nu} = 4k
 \label{extrinsic2}
\end{eqnarray}
The above expression of $Q$ (trace of the extrinsic curvature) leads to the boundary term of the action as,
\begin{eqnarray}
 S_{b}&=&\int d^4x\bigg[\sqrt{-g_h}\bigg(e^{-\frac{5}{2\sqrt{3}}\kappa v_h} \big(V_h+\frac{4k}{\kappa^2}\big) 
 - \frac{ab}{2\sqrt{3}k\kappa^2}e^{b\kappa v_h}\bigg)\nonumber\\
 &+&\sqrt{-g_v}\bigg(e^{-\frac{5}{2\sqrt{3}}\kappa v_v} \big(V_v+\frac{4k}{\kappa^2}\big) 
 - \frac{ab}{2\sqrt{3}k\kappa^2}e^{b\kappa v_v}\bigg)\bigg]\nonumber\\
 &=&\int d^4x\bigg[\sqrt{-g_h} V_h^{eff} + \sqrt{-g_v} V_v^{eff}\bigg]
 \label{boundary_action}
\end{eqnarray}

where we use the explicit solution of $\Psi(\phi)$ (see eqn.(\ref{sol.scalar.field})) with $v_h=\Psi(0)$, $v_v=\Psi(\pi)$. Further 
\begin{eqnarray}
 V_h^{eff} = e^{-\frac{5}{2\sqrt{3}}\kappa v_h} \bigg(V_h+\frac{4k}{\kappa^2}\bigg) - \frac{ab}{2\sqrt{3}k\kappa^2}e^{b\kappa v_h}
 \label{effective_tension1}
\end{eqnarray}
and
\begin{eqnarray}
 V_v^{eff} = e^{-\frac{5}{2\sqrt{3}}\kappa v_v} \bigg(V_v+\frac{4k}{\kappa^2}\bigg) - \frac{ab}{2\sqrt{3}k\kappa^2}e^{b\kappa v_v}
 \label{effective_tension2}
\end{eqnarray}

with $g_h$, $g_v$ are the determinants of the induced metric on hidden, visible brane respectively. It may be observed that the 
boundary terms emerging from the total derivative of $\Psi$ and the Gibbons-Hawking terms modify the brane tensions of the 
respective branes to produce the effective brane tensions as $V_h^{eff}$ and $V_v^{eff}$.

Plugging back the solution of $\Psi(\phi)$ (eqn.(\ref{sol.scalar.field})) into scalar field action and 
integrating over $\phi$ yields an effective modulus potential having the form as,
\begin{eqnarray}
 &V&_{eff}(r_c) = \frac{1}{b^2\kappa^2}\bigg[\frac{e^{-4kr_c\pi}}{\pi r_c-y_0} + \frac{1}{y_0}\bigg] 
 - \frac{\Lambda}{2k}\bigg[1- e^{-4kr_c\pi}\bigg]\nonumber\\
 &-&\frac{4k}{b^2\kappa^2}e^{-4ky_0}\bigg(Ei[4k(y_0-\pi r_c)] - Ei[4ky_0]\bigg)
 \label{mod.potential}
\end{eqnarray}
where 'Ei' denotes the exponential integral function.\\
It may be observed that  the scalar field degrees of 
freedom is related to the curvature as,
\begin{equation}
 \Psi(\phi) = \frac{2}{\sqrt{3}\kappa}\ln\big[1 + n\alpha R^{n-1}\big]
 \label{scalar and curvature}
\end{equation}
From the above expression, we can relate the boundary values of the scalar field 
(i.e $\Psi(0)=v_h$) with the  Ricci scalar as,
\begin{equation}
 v_h = \frac{2}{\sqrt{3}\kappa} \ln\big[1 + n\alpha R^{n-1}(0)\big]
 \label{relation}
\end{equation}
where $R(0)$ is the value of the curvature on Planck brane. Thus the parameters that are used in 
the scalar-tensor theory  are actually related to the 
parameters of the original $F(R)$ theory.\\
Furthermore the various components of stress tensor of the scalar field $\Psi$ can be written as,

 \begin{equation}
  T_{\mu\nu}(\Psi) = -\frac{1}{2}\eta_{\mu\nu}e^{-2kr_c|\phi|} \bigg[|\Lambda| - \frac{1}{2\kappa^2b^2} \frac{1}{(r_c\phi-y_0)^2}\bigg]
  \nonumber\\
 \end{equation}
 and 
 \begin{equation}
  T_{\phi\phi}(\Psi) = \frac{1}{2}r_c^2 \bigg[|\Lambda| - \frac{3}{2\kappa^2b^2} \frac{1}{(r_c\phi-y_0)^2}\bigg]
  \nonumber\\
 \end{equation}
 
 where we use the solution of $\Psi(\phi)$ obtained in eqn.(\ref{sol.scalar.field}). These above expressions of $T_{MN}(\Psi)$ lead to 
 the ratio of corresponding component of stress tensor between bulk scalar field and bulk cosmological constant as,
 \begin{eqnarray}
  \bigg(\frac{T_{\mu\nu}(\Psi)}{T_{\mu\nu}(\Lambda)}\bigg)_{max} = 1 - \frac{a^2b^2}{32\kappa^2k^2|\Lambda|} e^{2b\kappa v_h}
  \nonumber
 \end{eqnarray}
 and
 \begin{eqnarray}
  \bigg(\frac{T_{\phi\phi}(\Psi)}{T_{\phi\phi}(\Lambda)}\bigg)_{max} = 1 - \frac{3a^2b^2}{32\kappa^2k^2|\Lambda|} e^{2b\kappa v_h}
  \nonumber
 \end{eqnarray} 
 
 where $T_{\phi\phi}(\Lambda)$ and $T_{\mu\nu}(\Lambda)$ are different components of stress tensor for the bulk cosmological 
 constant. It may be observed that for $e^{b\kappa v_h}<\frac{\kappa k\sqrt{|\Lambda|}}{ab}$, the stress tensor 
 for the scalar field ($\Psi$) is less than that for the bulk cosmological constant ($\Lambda$) for entire range of extra dimensional 
 coordinate (i.e $0<\phi<\pi$). This condition allows us to neglect the back-reaction of the scalar field 
 in comparison to bulk cosmological constant.\\
 To introduce the radion field we replace $r_c \rightarrow \tilde{T}(t)$ \cite{GW_radion}, where $\tilde{T}(t)$ is the fluctuation 
of the modulus around its vev and is known as radion field. 
Here, for simplicity 
we assume \cite{GW_radion} that this new field depends only on $t$. The corresponding metric ansatz is,
\begin{equation}
 d\tilde{s}^2 = e^{- 2 k\tilde{T}(t)|\phi|} \eta_{\mu\nu} dx^{\mu} dx^{\nu} - \tilde{T}^2(t)^2d\phi^2
 \label{grav.sol2.ST}
\end{equation} 
Recall that the quantities in tilde are reserved for ST theory. As mentioned earlier, the bulk scalar field 
$\Psi$ fulfills the requirement for generating the radion potential.\\
With the metric in eqn.(\ref{grav.sol2.ST}), the five dimensional 
Einstein-Hilbert part of the action yields the kinetic part of the radion field in the four dimensional effective action as \cite{GW_radion},
\begin{equation}
 S_{kin}[\tilde{T}] = \frac{12M^3}{k} \int d^4x \partial_{\mu}(e^{-k\pi \tilde{T}(t)})\partial^{\mu}(e^{-k\pi \tilde{T}(t)})
 \nonumber\\
\end{equation}
As we see that $\tilde{T}(t)$ is not canonical and thus we redefine the field by the following transformation,
\begin{equation}
 \tilde{T}(t) \longrightarrow \tilde{T}_{can}(t) = \sqrt{\frac{24M^3}{k}}e^{-k\pi \tilde{T}(t)}
 \label{trans.ST}
\end{equation}

Correspondingly the radion potential is obtained from eqn.(\ref{mod.potential}) by replacing $r_c$ by $\tilde{T}(t)$ i.e.
 \begin{eqnarray}
 &V&_{ST}(\tilde{T}) = \frac{1}{b^2\kappa^2}\bigg[\frac{e^{-4k\pi \tilde{T}(t)}}{\pi \tilde{T}(t)-y_0} + \frac{1}{y_0}\bigg] 
 - \frac{\Lambda}{2k}\bigg[1- e^{-4k\pi\tilde{T}}\bigg]\nonumber\\
 &-&\frac{4k}{b^2\kappa^2}e^{-4ky_0}\bigg(Ei[4k(y_0-\pi \tilde{T})] - Ei[4ky_0]\bigg)
 \label{radion.potential.ST}
\end{eqnarray}

In terms of $\tilde{T}_{can}$, 
the Lagrangian of radion field becomes
\begin{equation}
 L[\tilde{T}_{can}] = \bigg[\frac{1}{2} \tilde{\dot{T}}_{can}^2 - V_{ST}(\tilde{T}_{can})\bigg]
 \nonumber\\
\end{equation}
which is the same Lagrangian for a particle moving in a potential $V_{ST}$.\\

\begin{figure}[!h]
\begin{center}
 \centering
 \includegraphics[width=3.0in,height=2.0in]{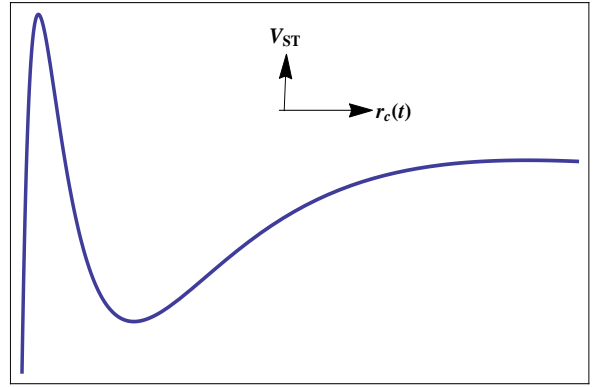}
 \caption{$V_{ST}$ vs $r_c(t)\big(=\tilde{T}(t)\big)$ for $a=1$, $b=\frac{\sqrt{2}}{3}$, $k=M=1$, $\Lambda=-1$, $\kappa v_h =0.01$}
 \label{plot potential1}
\end{center}
\end{figure}

The potential $V_{ST}$ has a minimum at
\begin{eqnarray}
 <\pi\tilde{T}>_{+}&=&<\pi r_c>_{+}\nonumber\\
 &=&y_0 - \frac{2k}{b^2\kappa^2\Lambda}\bigg[\sqrt{1-\frac{b^2\kappa^2\Lambda}{8k^2}} + 1\bigg]
 \label{min st}
\end{eqnarray}
and a maxima at
\begin{eqnarray}
 <\pi\tilde{T}>_{-}&=&<\pi r_c>_{-}\nonumber\\
 &=&y_0 - \frac{2k}{b^2\kappa^2\Lambda}\bigg[\sqrt{1-\frac{b^2\kappa^2\Lambda}{8k^2}} - 1\bigg]
 \label{max st}
\end{eqnarray}
respectively. Recall $y_0 = \frac{4k}{ab^2}e^{-b\kappa v_h}$, $a$ and $b$ are given by eqn.(\ref{a and b}). 
Moreover, eqn.(\ref{radion.potential.ST}) clearly indicates that $V_{ST}(\tilde{T})$ becomes zero at 
$\tilde{T}=0$ and reaches a constant value $\big(= \frac{1}{b^2\kappa^2y_0} - \frac{\Lambda}{2k} + \frac{4k}{b^2\kappa^2}
e^{-4ky_0}Ei[4ky_0]\big)$ asymptotically at large $\tilde{T}$. In figure(\ref{plot potential1}), we give the plot between 
$V_{ST}(\tilde{T})$ versus $\tilde{T}(t)$.

Consequently, using the form of radion potential in eqn.(\ref{radion.potential.ST}) with the 
transformation given in eqn.(\ref{trans.ST}), one 
arrives at the following mass squared of radion field in scalar-tensor theory given as,
\begin{eqnarray}
 \tilde{m}^2_{rad}(ST)&=&e^{-2k\pi <r_c>_{+}} \bigg(\frac{b^2\kappa^2\Lambda^2}{12M^3k^2}\bigg)*\nonumber\\
 &\bigg[&\frac{\sqrt{1-\frac{b^2\kappa^2\Lambda}{8k^2}}}{\big(1+\sqrt{1-\frac{b^2\kappa^2\Lambda}{8k^2}}\big)^2}\bigg]
 \label{radion.mass.ST}
\end{eqnarray}

According to Goldberger-Wise (GW) stabilization mechanism \cite{GW,GW_radion}, the modulus is stabilized at that separation 
for which the  effective radion potential becomes minimum. Therefore, in the present context, the stable value of interbrane 
separation is given by $r_c=<r_c>_+$, which is determined in eqn.(\ref{min st}). But due to quantum 
fluctuation, the radion field has a non zero probability to tunnel from $r_c=<r_c>_+$ to $r_c=0$, which 
in turn makes the brane configuration unstable. So it is worthwhile to calculate the quantum tunneling 
for radion field from $r_c=<r_c>_+$ to $r_c=0$. In order to do so, the radion potential is approximately 
considered as a rectangle barrier having width ($w$) $=<r_c>_+$ and height ($h$) $=V_{eff}(<r_c>_-)$ 
respectively. For such a potential barrier, the tunneling probability ($P_{ST}$) is given by,
\begin{eqnarray}
 \frac{1}{P_{ST}}&=&1 + \bigg(\frac{V_{ST}(<r_c>_-)}{\Delta V_{ST}}\bigg) 
 \sinh^2\bigg[\pi\sqrt{\frac{2\tilde{m}_{rad}\Delta V_{ST}}{M^3}}*\nonumber\\
 &<&r_c>_+ e^{-\frac{3}{2}k\pi <r_c>_+}\bigg]
 \label{probability st}
\end{eqnarray}

where $\tilde{m}_{rad}$ is the mass of radion field, determined in eqn.(\ref{radion.mass.ST}) and $\Delta V_{ST} = 
V_{ST}(<r_c>_-) - V_{ST}(<r_c>_+)$, which can be easily calculated from the expression of radion potential.\\
Obviously, $P_{ST}$ depends on the parameters $a$ and $b$. For $a=1$ (in Planckian unit), we give the 
plot between $P_{ST}$ versus $b$ (see figure(\ref{plot probability1})) :

\begin{figure}[!h]
\begin{center}
 \centering
 \includegraphics[width=3.5in,height=2.5in]{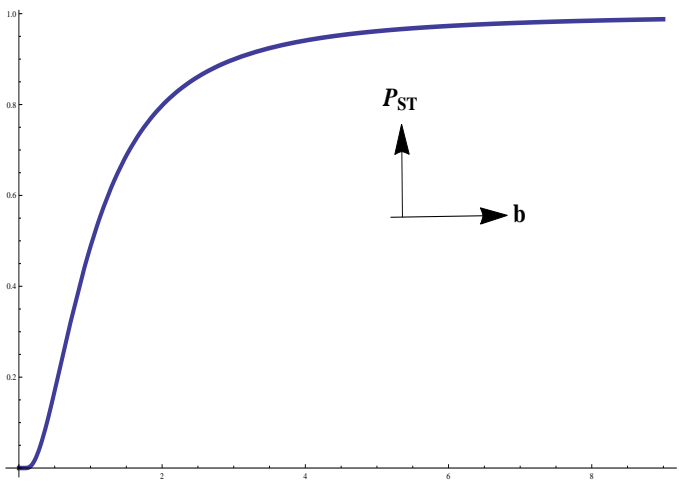}
 \caption{$P_{ST}$ vs $b$ for $a=1$, $k=M=1$, $\Lambda=-1$, $\kappa v_h =0.01$}
 \label{plot probability1}
\end{center}
\end{figure}

Figure(\ref{plot probability1}) clearly depicts that the tunneling probability increases with the parameter $b$ 
and asymptotically reaches to unity at large $b$. It is 
expected, because with increasing value of $b$, the height 
as well as the width $\big($ both are $\varpropto \frac{1}{b^2} \big)$ of the potential barrier decreases and 
as a consequence, $P_{ST}$ increases. Moreover, $P_{ST}$ becomes zero as $b$ tends to zero, because for $b \rightarrow 0$, 
the height of the potential barrier goes to infinite and as a result, $P_{ST} = 0$. This character of global minimum (as $b$ tends to zero) 
actually overlaps with the Goldberger-Wise result \cite{GW,GW_radion}.\\
However, resolution of the gauge hierarchy problem requires 
$k\pi<r_c>_+ = 36$, which in turn makes $b \simeq \frac{1}{3\sqrt{3}}$ (for $a=1$). With these values of $a$ and $b$, 
$P_{ST}$ becomes drastically suppressed and comes as $\sim 10^{-32}$. This small value of tunneling probability 
guarantees the stability of the interbrane separation (and hence of the radion field) at $<r_c>_+$. 
Thus it can be argued that the smallness of tunneling 
probability is intimately connected with the requirement of resolving the gauge hierarchy problem. Further it may be mentioned that these values 
of $a$ and $b$ are consistent with the condition $e^{b\kappa v_h}<\frac{\kappa k\sqrt{|\Lambda|}}{ab}$, necessary for 
neglecting the backreaction of the scalar field $\Psi$ in the background spacetime (as mentioned earlier).\\
Now we turn our focus on radion potential as well as on radion tunneling probability for 
the original $F(R)$ model (eqn.(\ref{actionF(R)})).

\section{Radion potential and tunneling probability in F(R) model}
Recall that the original higher curvature $F(R)$ model is described by the action given in eqn.(\ref{actionF(R)}). 
Solutions of metric ($G_{MN}$) for this $F(R)$ model can be extracted from the solutions of corresponding scalar-tensor 
theory (eqn.(\ref{grav.sol1.ST}) and eqn.(\ref{sol.scalar.field})) with the help of eqn.(\ref{conformal}). Thus 
the line element (in the bulk) in $F(R)$ model turns out to be
\begin{equation}
 ds^2 = e^{-\frac{\kappa}{\sqrt{3}}\Psi(\phi)} \bigg[e^{- 2 kr_c|\phi|} \eta_{\mu\nu} dx^{\mu} dx^{\nu} - r_c^2d\phi^2\bigg]
 \label{grav.sol1.F(R)}
\end{equation}
where $\Psi(\phi)$ is given in eqn.(\ref{sol.scalar.field}).\\
At this point, we need to verify whether the above solution of $G_{MN}$ (in eqn.(\ref{grav.sol1.F(R)})) satisfies the field equations 
of the original $F(R)$ theory. The five dimensional gravitational field equation for $F(R)$ theory is given by,
\begin{eqnarray}
 \frac{1}{2}G_{MN}F(R) - R_{MN}F'(R)&-&G_{MN}\Box F'(R)\nonumber\\ 
 &+&\nabla_{M}\nabla_{N}F'(R) = 0
 \label{field1}
\end{eqnarray}

In the present context, we take the form of $F(R)$ as $F(R) = R + \alpha R^n$ and thus the above field equation is simplified to the form :
\begin{eqnarray}
 &\frac{1}{2}&G_{MN}R - R_{MN} + \frac{\alpha}{2}G_{MN}R^n - n\alpha R^{n-1}R_{MN}\nonumber\\ 
 &-&n\alpha G_{MN}\Box R^{n-1} + n\alpha \nabla_{M}\nabla_{N}R^{n-1} = 0
 \label{field2}
\end{eqnarray}

It may be shown that the solution of $G_{MN}$ in eqn.(\ref{grav.sol1.F(R)}) satisfies the above field equation to the leading order 
of $\kappa v_h$. It may be recalled that the equivalence of the chosen $F(R)$ model was transformed to the potential 
of the scalar-tensor model in the leading order of $\kappa v_h$. Thus it 
guarantees the validity of the solution of spacetime metric (i.e. $G_{MN}$) in the original $F(R)$ theory.\\
However this solution 
of $G_{MN}$ immediately leads to the separation between hidden ($\phi=0$) and visible ($\phi=\pi$) branes along 
the path of constant $x^{\mu}$ as follows :
\begin{equation}
 \pi d = r_c \int_{0}^{\pi} d\phi e^{-\frac{\kappa}{2\sqrt{3}}\Phi(\phi)} 
 \nonumber\\
\end{equation}
where $d$ is the interbrane separation in $F(R)$ model. 
Using the explicit functional form of $\Psi(\phi)$ (eqn.(\ref{sol.scalar.field})), above equation can be integrated 
and simplified to the following one,
\begin{equation}
 k\pi d = k\pi r_c \bigg[1 - \frac{\kappa}{2\sqrt{3}}v_h + \frac{ab^2}{16\sqrt{3}k} \pi r_c\bigg]
 \label{brane.separation.F(R)}
\end{equation}
where the sub-leading terms of $\kappa\Psi$ are neglected. $r_c$ is the modulus in the corresponding ST theory 
and has a stable value at $<r_c>_+$, which is shown in the previous section 
(see eqn.(\ref{min st})). So, it can be argued that due to the stabilization of ST theory, the modulus $d$ 
in $F(R)$ model has also a stable value at,
\begin{eqnarray}
 k\pi<d>_+&=&k\pi<r_c>_+ \bigg[1 - \frac{\kappa}{2\sqrt{3}}v_h + \frac{1}{4\sqrt{3}}e^{-b\kappa v_h}\nonumber\\
 &-&\frac{a}{\big(8\sqrt{3}\kappa^2\Lambda\big)} \bigg(1+\sqrt{1-\frac{b^2\kappa^2\Lambda}{8k^2}}\bigg)\bigg]
 \label{stabilized.modulus.F(R)}
\end{eqnarray}

A fluctuation of branes around the stable configuration is now considered. This fluctuation 
can be taken as a field ($T(t)$) and for simplicity, this new field is assumed to be the 
function of $t$. The metric takes the following form,
\begin{eqnarray}
 ds^2 = e^{-\frac{\kappa}{\sqrt{3}}\Psi(t,\phi)} &[&e^{- 2 kT(t)|\phi|} 
 \eta_{\mu\nu} dx^{\mu} dx^{\nu}\nonumber\\ 
 &-&T(t)^2d\phi^2]
 \label{grav.sol2.F(R)}
\end{eqnarray}
From the angle of four dimensional effective theory, $T(t)$ is known as radion field. Moreover 
$\Psi(t,\phi)$ is obtained from eqn.(\ref{sol.scalar.field}) by replacing $r_c$ to $T(t)$. 
Plugging back the solution (see eqn.(\ref{grav.sol2.F(R)})) into five dimensional 
action and integrating over $\phi$ generates a kinetic as well as a potential part for the radion field $T(t)$. Kinetic part 
comes as
\begin{eqnarray}
 S_{kin}[T] = \frac{1}{2} f^2\int d^4x \partial_{\mu}(e^{-k\pi T(x)})\partial^{\mu}(e^{-k\pi T(x)}) 
 \nonumber\\
\end{eqnarray}

where the factor $f$ has the following form:
\begin{eqnarray}
 f&=&\sqrt{\frac{24M^3}{k}}\nonumber\\
 &\bigg[&1 + \bigg(\frac{40}{\sqrt{3}}\bigg(\frac{\sqrt{3}ak^2\kappa^{8/3}}{4b^2}\bigg)^{\frac{3}{2+2\sqrt{3}b}}
 \kappa v_h\bigg)^{\frac{5+2\sqrt{3}b}{4+4\sqrt{3}b}}\bigg]^{1/2}
 \label{factor f}
\end{eqnarray}

\begin{figure}[!h]
\begin{center}
 \centering
 \includegraphics[width=3.0in,height=2.0in]{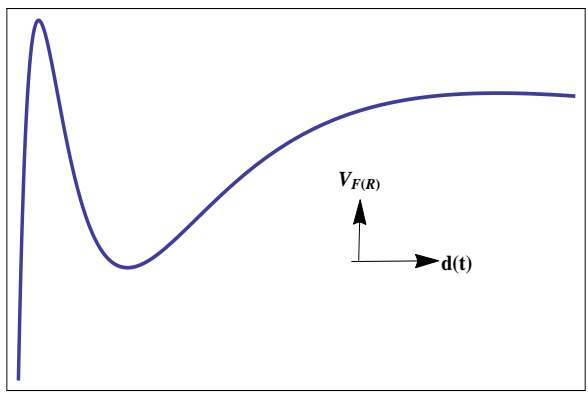}
 \caption{$V_{F(R)}$ vs $d(t)\big(=T(t)\big)$ for $a=1$, $b=\frac{\sqrt{2}}{3}$, $k=M=1$, $\Lambda=-1$, $\kappa v_h =0.01$}
 \label{plot potential2}
\end{center}
\end{figure}

Due to the appearance of $f$, $T(x)$ is not canonical and in order to make it canonical, we 
redefine the field as 
\begin{equation}
 T(x) \longrightarrow T_{can}(x) = f e^{-k\pi T(x)}
 \label{trans.F(R)}
\end{equation} 
For $a \rightarrow 0$, the action contains only the linear term in Ricci scalar 
and the factor $f$ goes to $\sqrt{24M^3/k}$ which agrees with \cite{GW_radion}.\\
Finally the potential part of radion field is as follows,
\begin{eqnarray}
 V_{F(R)}(T_{can})&=&\frac{20}{\sqrt{3}} \bigg(\frac{f^4k^{\frac{5+2\sqrt{3}b}{1+\sqrt{3}b}}}{M^6}\bigg) 
 \bigg(\frac{\sqrt{3}a\kappa^{8/3}}{4b^2}\bigg)^{\frac{3}{2+2\sqrt{3}b}}\nonumber\\ 
 &\bigg[&1 + \frac{4}{\sqrt{3}}\big(\kappa v_h\big)^{\frac{5+2\sqrt{3}b}{4+4\sqrt{3}b}}\bigg]*\nonumber\\
 &\bigg[&\frac{1}{b^2\kappa^2}\bigg(\frac{1}{y_0} - \frac{kT_{can}^4}{f^4\big(y_0-\ln{f}+\ln{T_{can}\big)}}\nonumber\\
 &-&\frac{b^2\kappa^2\Lambda}{2k}\bigg(1 - \frac{T_{can}^4}{f^4}\bigg)\bigg) - \frac{4k}{b^2\kappa^2}e^{-4ky_0}*\nonumber\\
 &\bigg(&Ei[4(ky_0-\ln{f}+\ln{T_{can}})]\nonumber\\ 
 &-&Ei[4ky_0]\bigg)\bigg]
 \label{potential.radion.F(R)}
\end{eqnarray}
It may be observed that $V_{F(R)}(T_{can})$ goes to zero as the parameter $a$ tends 
to zero. This is expected because for $a \rightarrow 0$, 
the action contains only the Einstein part ((recall 
from eqn.(\ref{a and b}) that the higher curvature parameter $\alpha$ is proportional to $a$)) 
which does not produce any 
potential term for the radion field \cite{GW_radion}. 
Thus for five dimensional warped geometric model, the radion potential is 
generated from the higher order curvature term $\alpha R^n$. Again the Lagrangian for the canonical radion takes the following form :
\begin{equation}
 L[T_{can}] = \bigg[\frac{1}{2} \dot{T}_{can}^2 - V_{F(R)}(T_{can})\bigg]
 \nonumber\\
\end{equation}
which matches with the Lagrangian for a point particle moving under a potential $V_{F(R)}$.\\
The radion potential in $F(R)$ model has also a minima and a maxima at $<T_{can}>_+$ and at $<T_{can}>_-$ 
respectively, where
\begin{equation}
 <T_{can}>_+ = f e^{-k\pi <d>_+}
 \label{min F(R)}
\end{equation}
and
\begin{eqnarray}
 <T_{can}>_- = f e^{-k\pi <d>_-}
 \label{max F(R)}
\end{eqnarray}

with $<d>_+$, $<d>_-$ have the following expressions :
\begin{eqnarray}
 <d>_{\pm}&=&<r_c>_{\pm} \bigg[1 - \frac{\kappa}{2\sqrt{3}}v_h + \frac{1}{4\sqrt{3}}e^{-b\kappa v_h}\nonumber\\
 &\mp&\frac{1}{8\sqrt{3}}\frac{a}{\kappa^2\Lambda} \bigg(\sqrt{1-\frac{b^2\kappa^2\Lambda}{8k^2}} \pm 1\bigg)\bigg]
 \label{brane_separation_F(R)}
 \end{eqnarray}
 where $<r_c>_{\pm}$ are determined in eqn.(\ref{min st}) and in eqn.(\ref{max st}) respectively. 
We emphasize that due to the presence 
of conformal factor connecting the two theories, the value of 
$<d>_{\pm}$ (in $F(R)$ model) is different from $<r_c>_{\pm}$ (in ST model). Finally 
the squared mass of radion field is as follows,
\begin{eqnarray}
 m_{rad}^2&=&e^{-2k\pi <d>_+} \big(\frac{5}{3\sqrt{3}}\big) 
 \bigg(\frac{\sqrt{3}a\kappa^{8/3}}{4b^2}\bigg)^{\frac{3}{2+2\sqrt{3}b}} 
 \bigg(\frac{k^{\frac{7+4\sqrt{3}b}{1+\sqrt{3}b}}}{M^2}\bigg)*\nonumber\\
 &\bigg[&1 + \frac{20}{\sqrt{3}} \bigg(\bigg(\frac{\sqrt{3}ak^2\kappa^{8/3}}{4b^2}\bigg)^{\frac{3}{2+2\sqrt{3}b}} 
 \kappa v_h\bigg)^{\frac{5+4\sqrt{3}b}{4+4\sqrt{3}b}}\bigg]\nonumber\\
 &\bigg[&\frac{\sqrt{1-\frac{b^2\kappa^2\Lambda}{8k^2}}}{\big[1 + \sqrt{1-\frac{b^2\kappa^2\Lambda}{8k^2}}\big]^2}\bigg]
 \label{radion.mass.F(R)}
\end{eqnarray}
 It is evident that mass of the radion field also goes to zero as $a \rightarrow 0$ (higher curvature parameter 
 $\alpha$ is proportional to $a$).\\
Using the form of $V_{F(R)}(T_{can})$ along with the transformation eqn.(\ref{trans.F(R)}), now 
we give the plot between radion potential and $T(t)$ (see figure(\ref{plot potential2})).

\begin{figure}[!h]
\begin{center}
 \centering
 \includegraphics[width=3.5in,height=2.5in]{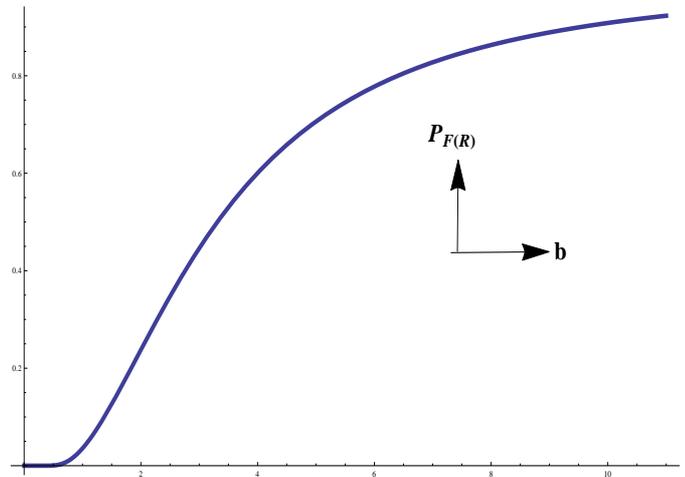}
 \caption{$P_{F(R)}$ vs $b$ for $a=1$, $k=M=1$, $\Lambda=-1$, $\kappa v_h =0.01$}
 \label{plot probability2}
\end{center}
\end{figure}

Figure(\ref{plot potential2}) clearly depicts that the radion potential goes to zero at $T(x)=0$ and reaches 
a constant value asymptotically at large value of $T(x)$. Comparing figure(1) and figure(3), 
it is clear that the nature of radion potential does not change in comparison to that in ST theory. 
However, due to the conformal factor, the extremas of the potential are shifted in $F(R)$ model, 
which is clear from eqn.(\ref{brane.separation.F(R)}).\\
As per GW mechanism, the radion field is stabilized at $<d>_+$. But as mentioned earlier, due to quantum 
mechanical tunneling effect, there exists a non zero tunneling probability of the radion field from $d = <d>_+$ to 
$d = 0$. Again considering the radion potential as a rectangle barrier 
having width ($w = <d>_+$) and height ($h = V_{F(R)}(<d_->)$), we calculate the tunneling probability ($P_{F(R)}$) 
from $d = <d>_+$ to $d = 0$ and is given by,
\begin{eqnarray}
 \frac{1}{P_{F(R)}}&=&1 + \bigg(\frac{V_{F(R)}(<d>_-)}{\Delta V_{F(R)}}\bigg) 
 \sinh^2\bigg[\pi\sqrt{\frac{2{m}_{rad}\Delta V_{F(R)}}{M^3}}*\nonumber\\
 &<&d>_+ e^{-\frac{3}{2}k\pi <d>_+}\bigg]
 \label{probability F(R)}
\end{eqnarray}
where $m_{rad}$ is given in eqn.(\ref{radion.mass.F(R)}) and $\Delta V_{F(R)} = V_{F(R)}(<d_->) - V_{F(R)}(<d_+>)$. 
From eqn.(\ref{probability F(R)}), it is clear that $P_{F(R)}$ depends on both the parameters $a$ and $b$. Here we take 
$a = 1$ (in Planckian unit) and give the plot demonstrating the variation of $P_{F(R)}$ with respect to $b$ (see figure(\ref{plot probability2})).

Figure(\ref{plot probability2}) reveals that just as in ST theory, $P_{F(R)}$ (tunneling probability in F(R) model) increases with increasing value 
of $b$ and acquires the maximum value ($= 1$) asymptotically at large $b$. For $b\rightarrow \infty$, the higher curvature 
parameter ($\alpha$) goes to zero (see eqn.(\ref{a and b})) and the action reduces to Einstein-Hilbert action. This in turn 
makes the brane configuration unstable \cite{GW_radion} and as a consequence the tunneling probability becomes unity. As the parameter 
$b$ decreases, the effect of higher curvature term starts to contribute and as a result, the modulus is stabilized at a certain separation 
and hence the probability for tunneling becomes less than one. Furthermore for $b\rightarrow 0$, higher curvature parameter $\alpha\rightarrow \infty$, 
which in turn makes the height of the radion potential barrier infinity (height $\varpropto \frac{1}{b^2}$) and thus the potential 
acquires a global minimum. As a consequence, the tunneling probability tends to zero, which is shown in figure (\ref{plot probability2}). 
The character of global minimum actually mimics the result of Goldberger and Wise \cite{GW}. It is expected 
because for $b\rightarrow 0$, the bulk scalar 
potential in the present context ($U(\Psi)$) becomes quadratic (all the other terms are proportional to higher power of $b$ and can 
be neglected) as same as the potential considered in \cite{GW}.\\
Finally we examine whether the solution of gauge hierarchy problem in F(R) model leads to a small value of the tunneling probability 
or not. We find that the resolution of gauge hierarchy problem requires $k\pi <d>_+ = 36$, which in turn makes 
$b = \frac{\sqrt{2}}{3}$. For this value of $b$, $P_{F(R)}$ is highly suppressed and takes the value of $\sim 10^{-32}$. Therefore, 
in original $F(R)$ theory, the requirement for solving the gauge hierarchy problem is correlated with the smallness of radion 
tunneling probability (a similar analysis is also obtained in ST theory as discussed in section [IV]).

\section{Conclusion}
In this work, we consider a five dimensional compactified warped geometry model with two 3-branes embedded within the spacetime. 
Due to large curvature ($\sim$ Planck scale), the bulk spacetime is governed by a higher curvature theory like $F(R) = R + \alpha R^{n}$. 
In this scenario, we determine the radion potential from the scalar degrees of freedom of higher curvature gravity and 
investigate the possibility of tunneling for the radion field. Our findings and implications are as follows :

\begin{itemize}
 \item Due to the presence of higher curvature gravity in the bulk, a potential term for the radion field is generated, 
 as shown in eqn.(\ref{potential.radion.F(R)}). This is in sharp contrast to a model with only 
 Einstein term in the bulk where the modulus potential can not be generated 
 without incorporating any external degrees of freedom such as a scalar field. However for the higher curvature gravity model, 
 this additional degree of freedom originates naturally from the higher curvature term. It may also be noted that 
 the radion potential goes to zero as the higher curvature parameter $\alpha \rightarrow 0$.
 
 \item The radion potential ($V_{F(R)}$) has a minimum ($<d>_+$) and a maximum ($<d>_-$) respectively where the height between minimum 
 and maximum of the potential depends on both the parameters $\alpha$ and $n$. Moreover, $V_{F(R)}$ becomes zero at $T(x) = 0$ 
 ($T(x)$ is the radion field) and reaches a constant value asymptotically at large $T(x)$, as depicted in figure(\ref{plot potential2}).
 
 \item According to GW mechanism, the modulus is stabilized at $<d>_+$. But due to quantum mechanical effect, there exists a possibility of 
 tunneling for the radion field from $d = <d>_+$ to $d = 0$, which in turn makes the aforementioned brane configuration unstable. We 
 calculate this tunneling probability ($P_{F(R)}$) which depends on the parameters $a$ and $b$ ($a$ and $b$ can be written in terms 
 of $\alpha$ and $n$, see eqn.(\ref{a and b})). For a certain choice of $a$, $P_{F(R)}$ increases with increasing value of $b$, as demonstrated 
 in figure(\ref{plot probability2}). It may be observed that this behaviour of $P_{F(R)}$ with the parameter $b$ is expected, because 
 the height of the potential barrier decreases as $b$ increases and as a result, $P_{F(R)}$ increases. Finally we find that 
 the solution of gauge hierarchy problem requires $k\pi<d>_+ = 36$, which in turn highly suppresses the tunneling probability and as 
 a consequence, $P_{F(R)}$ comes as $\sim 10^{-32}$. This small value of the tunneling probability guarantees the stability of 
 interbrane separation at $<d>_+$. Therefore, it can be argued that the smallness of tunneling probability is interrelated 
 with the requirement of solving the gauge hierarchy problem.
\end{itemize}

\end{document}